\documentclass[aps,prl,twocolumn,showpacs,groupedaddress,amsmath,amssymb]{revtex4-1}
\usepackage{mathrsfs}

\usepackage{graphicx}
\usepackage{bm}
\usepackage{float}

\begin{document}

\title{Quantum superposition of localized and delocalized phases of photons}

\author{Chun-Wang Wu}
  \email{Electronic address: cwwu@nudt.edu.cn}
\author{Ming Gao}
\author{Zhi-Jiao Deng}
\author{Hong-Yi Dai}
\author{Ping-Xing Chen}
\author{Cheng-Zu Li}
\affiliation{College of Science, National University of Defense
 Technology, Changsha 410073, People's Republic of China}

\date{\today}

\begin{abstract}
 Based on a variant of 2-site Jaynes-Cummings-Hubbard model, which is constructed using superconducting circuits, we propose a
 method to coherently superpose the localized and delocalized phases of photons. In our model, two nonlinear superconducting
 stripline resonators are coupled by an interfacial circuit composed of parallel combination of a superconducting qubit and a
 capacitor, which plays the role of a quantum knob for the photon hopping rate: with the knob qubit in its ground/excited state,
 the injected photons tend to be localized/delocalized in the resonators. We show that, by applying a microwave field with
 appropriate frequency on the knob qubit, we could demonstrate Rabi oscillation between photonic localized phase and delocalized phase.
 Furthermore, this set-up offers advantages (e.\,g.\,infinite on/off ratio) over other proposals for the realization of scalable
 quantum computation with superconducting qubits.
\end{abstract}


\maketitle

 \emph{Introduction.---}Coupled arrays of nonlinear resonators have recently been shown to be suitable candidates for exploring
 quantum many-body phenomena of light \cite{reve1}. So far, various strongly correlated effects and exotic phases have been studied
 using these artificial structures. Examples include effective photon-photon repulsion \cite{reve2}, the Mott insulator-superfluid
 quantum phase transition of light \cite{reve3}, photonic Josephson effect \cite{reve4}, and time-reversal-symmetry breaking \cite{reve5}.
 Compared to other strongly interacting many-particle systems, like Josephson junction arrays \cite{reve6} and optical lattices \cite{reve7},
 coupled resonator arrays have the advantage of accessing individual sites experimentally.

 Much recent work has focused on using polaritons in an array of coupled nonlinear resonators, described by the Jaynes-Cummings-Hubbard
 model (JCHM), to simulate the famous Bose-Hubbard model \cite{reve3,reve8}. In a coupled resonator array, the strong atom-photon
 coupling inside the resonator leads to an effective polariton repulsion, and the photon hopping
 between neighboring resonators favors delocalization of the polaritons. As the ratio of the hopping term relative to the on-site
 repulsion is varied through the quantum critical point, the ground state of the system undergoes a transition from a product of
 localized states of definite polariton number to a delocalized state with large fluctuations in the polariton number per site.

 According to the linear superposition principle of quantum mechanics, any linear combination of two allowed states of a system
 is also an allowed state. The quantum phases are some special states of many-particle systems. It is thus natural to
 ask, can we produce the superposition involving distinct quantum phases at will? The purpose of this letter
 is to explore the possibility of superposing the localized and delocalized phases of polaritons in a coupled resonator
 array. In order to keep the complexity of the system to a manageable level, we mainly consider the simplest possible case, i.\,e.\,two
 coupled resonators containing a total of two polaritons. In this case, it should be understood that in our usage the term ``phase'' refers
 to a certain state of a small finite system, not true phase in the thermodynamic sense.

 In our proposed model, two nonlinear stripline resonators are coupled by an interfacial circuit playing the role of a
 quantum knob, which is composed of parallel combination of a superconducting qubit and a capacitor. Corresponding to the two basis states
 of the knob qubit, polaritons in the resonators are governed by different effective Hamiltonians, which favor localization and delocalization
 of the polaritons, respectively. We show that, by applying spectroscopic techniques to the knob qubit, we can demonstrate Rabi oscillation
 between photonic localized phase and delocalized phase. This opens up a way to make photons or polaritons
 enter novel quantum states and enables new investigations of many-body physics in coupled resonator arrays. In addition, this architecture
 can be used to solve the annoying on/off ratio problem of conventional proposals for scalable quantum computation with superconducting qubits.

 \begin{figure}[!b]
 \includegraphics*[scale=0.65]{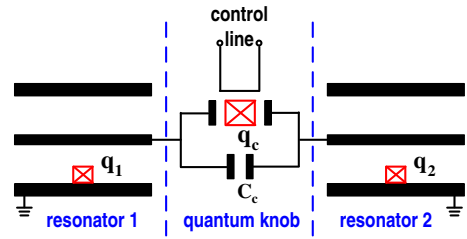}
 \caption{\label{fig1}(Color online) Two microwave stripline resonators, each containing a superconducting qubit, are coupled by a quantum
 knob composed of parallel combination of a qubit and a capacitor. The knob qubit is capacitively coupled to the resonators and can be driven
 by applying a microwave field on the control line. }
 \end{figure}

 \emph{Model and Hamiltonian.---}A sketch of our model is shown in Fig.\,1. Two microwave stripline resonators, each containing a superconducting
 qubit (e.\,g.\,phase \cite{reve9} or charge \cite{reve10} qubit), are coupled by an interfacial circuit composed of parallel combination of a
 qubit $q_{c}$ and a capacitor $C_{c}$, which plays the role of a quantum knob. The knob qubit $q_{c}$ is capacitively coupled to the resonators
 and can be driven by applying a microwave field on the control line. The capacitor $C_{c}$ leads to direct photon hopping between the resonators \cite{reve11}.
 Not considering the driving field now, the total system can be described by the Hamiltonian (assuming $\hbar=1$)
 \setlength{\abovedisplayskip}{5pt}
 \setlength{\belowdisplayskip}{5pt}
 \setlength{\abovedisplayshortskip}{5pt}
 \setlength{\belowdisplayshortskip}{5pt}
 \begin{equation}
 H=\sum_{i=1,2}H_{i}^{JC}+H^{c},
 \end{equation}
 which includes the Jaynes-Cummings (JC) interaction of the local resonator-qubit system
 \begin{equation}
 H_{i}^{JC}=\epsilon|e\rangle_{ii}\langle e|+wa_{i}^{\dag}a_{i}+g(\sigma_{i}^{+}a_{i}+\sigma_{i}^{-}a_{i}^{\dag})
 \end{equation}
 and the knob-mediated interaction between the resonators
 \begin{equation}
 H^{c}=\epsilon_{c}|e^{c}\rangle\langle e^{c}|+g_{c}(\sigma_{c}^{+}a_{1}+\sigma_{c}^{+}a_{2}+H.c.)+\kappa_{0}(a_{1}^{\dag}a_{2}+a_{1}a_{2}^{\dag}).
 \end{equation}
 Here, $\epsilon$, $\epsilon_{c}$ and $w$ are the resonance frequencies for $q_{i}$, $q_{c}$ and resonator $i$, respectively, $g$ ($g_{c}$) is the
 coupling strength between $q_{i}$ ($q_{c}$) and resonator $i$, and $\kappa_{0}$ is the fixed photon hopping rate between the resonators induced
 by the capacitor $C_{c}$. The states $|g\rangle_{i}$ ($|g^{c}\rangle$) and $|e\rangle_{i}$ ($|e^{c}\rangle$) are ground and excited states for
 $q_{i}$ ($q_{c}$), the operators $\sigma_{i}^{+}$ ($\sigma_{c}^{+}$) and $\sigma_{i}^{-}$ ($\sigma_{c}^{-}$) are qubit raising and lowering operators
 for $q_{i}$ ($q_{c}$), and $a_{i}^{\dag}$ ($a_{i}$) is the photon creation (annihilation) operator for resonator $i$.

 With $q_{c}$ working in the dispersive regime, i.\,e.\,$\Delta_{c}=\epsilon_{c}-w\gg g_{c}$, the real energy exchanges between $q_{c}$ and the resonators
 are largely suppressed. In this case, we can perform the unitary transformation $U=\exp[\frac{g_{c}}{\Delta_{c}}(a_{1}\sigma_{c}^{+}-a_{1}^{\dag}\sigma_{c}^{-}
 +a_{2}\sigma_{c}^{+}-a_{2}^{\dag}\sigma_{c}^{-})]$ and expand $UHU^{\dag}$ to second order in $\frac{g_{c}}{\Delta_{c}}$ to obtain the
 effective system Hamiltonian \cite{reve12}
 \begin{eqnarray}
 H^{eff}&=&\sum_{i=1,2}H_{i}^{JC}-\frac{g_{c}^{2}}{\Delta_{c}}(a_{1}^{\dag}a_{1}+a_{2}^{\dag}a_{2})|g^{c}\rangle\langle g^{c}|\nonumber\\
 &&+[\epsilon_{c}+\frac{2g_{c}^{2}}{\Delta_{c}}+\frac{g_{c}^{2}}{\Delta_{c}}(a_{1}^{\dag}a_{1}+a_{2}^{\dag}a_{2})]|e^{c}\rangle\langle e^{c}|\nonumber\\
 &&+(\kappa_{0}+\frac{g_{c}^{2}}{\Delta_{c}}\sigma_{c}^{z})(a_{1}^{\dag}a_{2}+a_{1}a_{2}^{\dag}),
 \end{eqnarray}
 where $\sigma_{c}^{z}=|e^{c}\rangle\langle e^{c}|-|g^{c}\rangle\langle g^{c}|$. The second and third terms of $H^{eff}$ represent the ac Stark shifted
 frequency of $q_{c}$, and the fourth term is the sum of the direct photon hopping induced by $C_{c}$ and the qubit-state-dependent photon hopping mediated
 by $q_{c}$. Choosing $\frac{g_{c}^{2}}{\Delta_{c}}=\kappa_{0}$, the parallel combination of $q_{c}$ and $C_{c}$ can function as a quantum knob \cite{reve13}:
 with $q_{c}$ in its ground/excited state, the photon hopping between the resonators will be switched off/on.

 Now, we verify the above results by means of numerical simulations. Initially, one photon is injected into resonator 1, with resonator 2 being empty. The
 parameters we choose are $g_{c}=g$, $\kappa_{0}=0.1g$, $\epsilon=45g$, $\epsilon_{c}=50g$, and $w=40g$, which yield $\frac{g_{c}^{2}}{\Delta_{c}}=\kappa_{0}$.
 In Fig.\,2, we plot the polariton numbers of resonator 1 (thick lines) and resonator 2 (thin lines) as functions of time, with the knob qubit in (a) $|g^{c}\rangle$,
 and (b) $|e^{c}\rangle$. The solid and dashed lines represent the dynamics governed by $H$ and $H^{eff}$, respectively. It is shown that, the evolution of the
 system can be described by $H^{eff}$ to a very good approximation, and we can really switch on and off the photon hopping by engineering the quantum state of $q_{c}$.

 \begin{figure}[!t]
 \includegraphics*[scale=0.58]{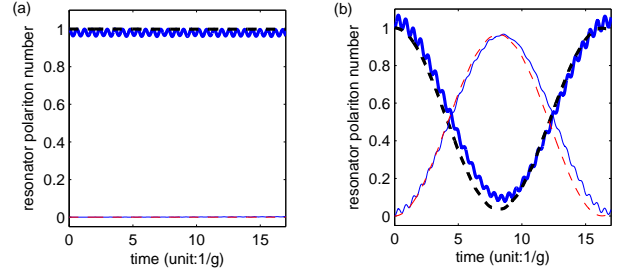}
 \caption{\label{fig2}(Color online) The polariton numbers of resonator 1 (thick lines) and resonator 2 (thin lines) as functions of time, with the knob qubit in (a) $|g^{c}\rangle$, and (b) $|e^{c}\rangle$. The solid and dashed lines represent the dynamics governed by $H$ and $H^{eff}$, respectively. }
 \end{figure}

 \emph{Eigenstates of the effective Hamiltonian $H^{eff}$.---}In the following, we will calculate the eigenstates of the system Hamiltonian and show that, by
 choosing appropriate parameters, our proposed 2-site JCHM can work in distinct regimes via only changing the internal state of $q_{c}$. For
 simplicity, we use the effective system Hamiltonian and restrict our analysis to the case of the resonators containing a total of two polaritons.

 If $\frac{g_{c}^{2}}{\Delta_{c}}=\kappa_{0}$ and $q_{c}$ is in $|g^{c}\rangle$, the photon hopping terms vanish and the system can be described by Hamiltonian
 $H^{eff}_{g}=\sum_{i=1,2}H_{i}^{JC}$. Here, we have written the ac Stark shifting terms into $H_{i}^{JC}$ and the resonator frequency $w$ is replaced by a
 shifted frequency $w^{'}=w-\frac{g_{c}^{2}}{\Delta_{c}}$. The local JC Hamiltonian $H_{i}^{JC}$ can be diagonalized in the basis of polariton states \cite{reve14}.
 Let $|n,g\rangle$ ($|n,e\rangle$) represent a resonator that contains $n$ photons and a single qubit in the ground (excited) state, then the upper and lower
 n-polariton states of resonator $i$ can be given by $|n+\rangle_{i}=\sin\theta_{n}|n-1,e\rangle_{i}+\cos\theta_{n}|n,g\rangle_{i}$, and $|n-\rangle_{i}=\cos\theta_{n}|n-1,e\rangle_{i}-\sin\theta_{n}|n,g\rangle_{i}$, respectively, where $\tan\theta_{n}=(\frac{\Delta}{2}+\sqrt{(\frac{\Delta}{2})^{2}+ng^{2}})/\sqrt{n}g$ and $\Delta=\epsilon-w^{'}$. Specially, the zero-polariton state $|0-\rangle_{i}=|0,g\rangle_{i}$. In order of increasing energy, the eight eigenstates of $H^{eff}_{g}$ are $\{|1-\rangle_{1}|1-\rangle_{2}$, $|2-\rangle_{1}|0-\rangle_{2}$, $|0-\rangle_{1}|2-\rangle_{2}$, $|1-\rangle_{1}|1+\rangle_{2}$, $|1+\rangle_{1}|1-\rangle_{2}$, $|2+\rangle_{1}|0-\rangle_{2}$, $|0-\rangle_{1}|2+\rangle_{2}$, $|1+\rangle_{1}|1+\rangle_{2}\}$. The state $|1-\rangle_{1}|1-\rangle_{2}$
 is exactly the ground state of 2-site JCHM in the localized regime.

 If $\frac{g_{c}^{2}}{\Delta_{c}}=\kappa_{0}$ and $q_{c}$ is in $|e^{c}\rangle$, the system can be described by $H^{eff}_{e}=\sum_{i=1,2}H_{i}^{JC}+2\kappa_{0}(a_{1}^{\dag}a_{2}+a_{1}a_{2}^{\dag})$. In this case, the resonator frequency $w$ is replaced by a shifted frequency
 $w^{'}=w+\frac{g_{c}^{2}}{\Delta_{c}}$ and the two resonators are coupled by a strength $J=2\kappa_{0}$. If $J$ is much smaller than the energy splitting between the upper polariton branch and lower polariton branch, then the mixing of different branches is negligible, and the lowest three eigenstates of $H^{eff}_{e}$ are linear combinations of $|1-\rangle_{1}|1-\rangle_{2}$, $|2-\rangle_{1}|0-\rangle_{2}$ and $|0-\rangle_{1}|2-\rangle_{2}$. If we further require that $J$ dominates over the effective repulsive energy $u_{r}$ between two polaritons of lower branch ($u_{r}$ equals the energy splitting between $|1-\rangle_{1}|1-\rangle_{2}$ and $|2-\rangle_{1}|0-\rangle_{2}$), then the coupled resonators work in the delocalized regime. In this case, the lowest three eigenstates of $H^{eff}_{e}$ can be approximated by the states $|\phi\rangle^{1}_{e}=\frac{1}{2}|2-\rangle_{1}|0-\rangle_{2}+\frac{1}{2}|0-\rangle_{1}|2-\rangle_{2}-\frac{\sqrt{2}}{2}|1-\rangle_{1}|1-\rangle_{2}$, $|\phi\rangle^{2}_{e}=\frac{\sqrt{2}}{2}|2-\rangle_{1}|0-\rangle_{2}-\frac{\sqrt{2}}{2}|0-\rangle_{1}|2-\rangle_{2}$, and $|\phi\rangle^{3}_{e}=\frac{1}{2}|2-\rangle_{1}|0-\rangle_{2}+\frac{1}{2}|0-\rangle_{1}|2-\rangle_{2}+\frac{\sqrt{2}}{2}|1-\rangle_{1}|1-\rangle_{2}$, respectively. These states are the eigenstates of 2-site JCHM in the large hopping limit. Note that, ``$J$ dominates over $u_{r}$'' does not
 mean $J\gg u_{r}$. Following the results of the pure Bose-Hubbard model, the quantum critical point of entering into delocalized regime is $J\simeq0.3u_{r}$ \cite{reve15}.

 \begin{figure}[!b]
 \includegraphics*[scale=0.5]{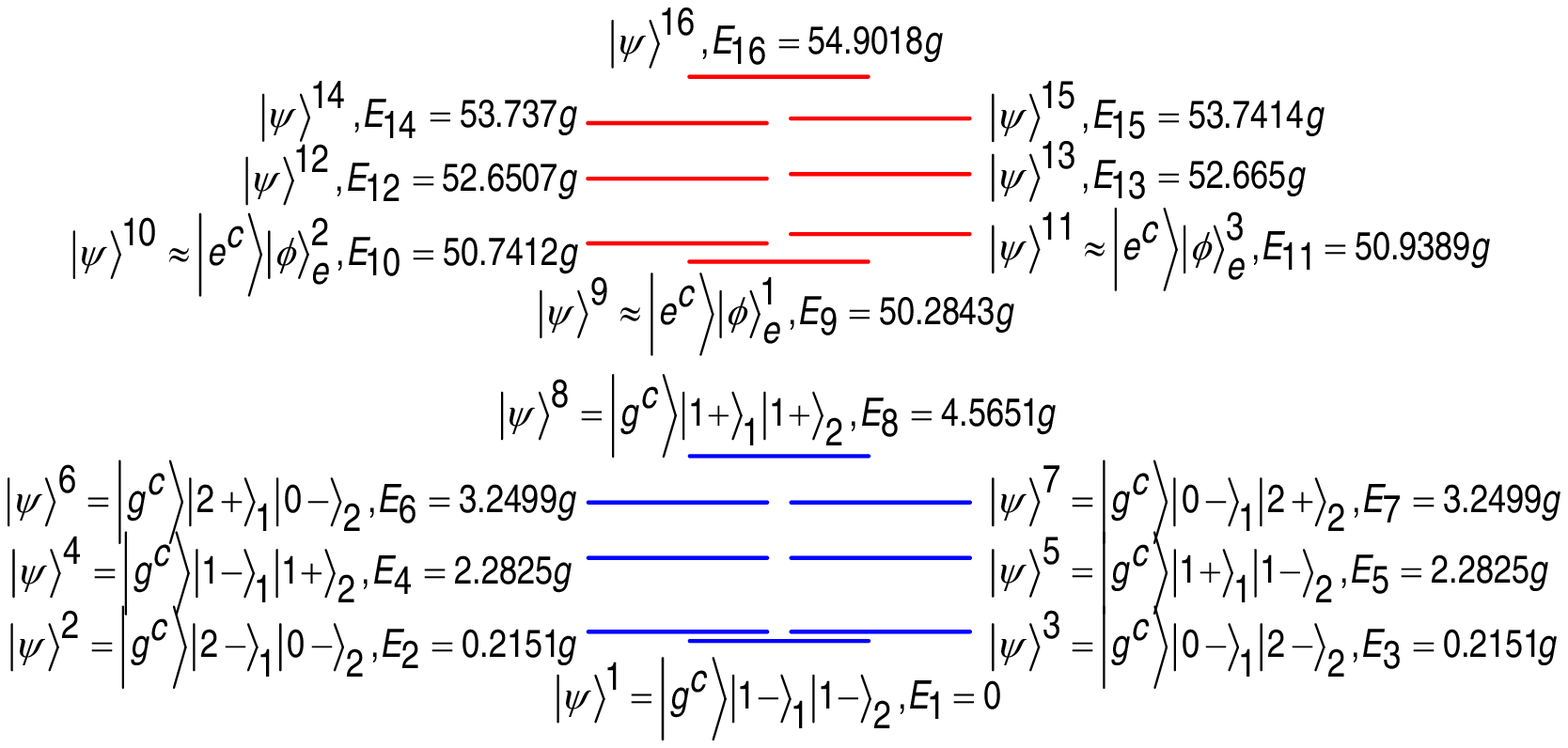}
 \caption{\label{fig3}(Color online) Spectrum of $H^{eff}$ with the resonators containing a total of two polaritons. The parameters we choose are $g_{c}=g$, $\kappa_{0}=0.1g$, $\epsilon=41g$, $\epsilon_{c}=50g$, and $w=40g$. }
 \end{figure}

 In Fig.\,3, we plot the spectrum of $H^{eff}$ with the resonators containing a total of two polaritons. The parameters we choose are $g_{c}=g$, $\kappa_{0}=0.1g$, $\epsilon=41g$, $\epsilon_{c}=50g$, and $w=40g$. The lower eight ($|\psi\rangle^{j}, 1\leq j\leq8$) and higher eight eigenstates ($|\psi\rangle^{k}, 9\leq k\leq16$) are two manifolds corresponding to $q_{c}$ in $|g^{c}\rangle$ and $|e^{c}\rangle$, respectively, $E_{l}$ ($1\leq l\leq16$) are the related sixteen eigenvalues, and the eigenstates $|\psi\rangle^{m}$ with $12\leq m\leq16$ all involve upper polariton states of the resonators. With these parameters, we have $u_{r}=0.259g$, $J=0.2g$, and $|^{9}\langle\psi|e^{c}\rangle|\phi\rangle_{e}^{1}|=0.980$, $|^{10}\langle\psi|e^{c}\rangle|\phi\rangle_{e}^{2}|=0.998$, $|^{11}\langle\psi|e^{c}\rangle|\phi\rangle_{e}^{3}|=0.979$. Therefore, by choosing these parameters, we can make this 2-site JCHM stay in localized and delocalized regime via engineering $q_{c}$ in $|g^{c}\rangle$ and $|e^{c}\rangle$, respectively.

 \emph{Quantum superposition of distinct quantum phases of photons.---}As shown above, our proposed system has the features of nonlinear spectrum and qubit-state-dependent light phases. These features allow us to demonstrate the Rabi oscillation between the localized phase and delocalized phase using a
 spectroscopic technique. Now, we apply a microwave field with frequency $w_{d}$ and strength $\Omega$ to the control line of $q_{c}$, then the total system Hamiltonian
 reads
 \begin{equation}
 H_{tot}=H+\Omega(\sigma_{c}^{+}e^{-iw_{d}t}+\sigma_{c}^{-}e^{iw_{d}t}).
 \end{equation}
 If $w_{d}$ is chosen to be equal to the energy splitting between $|\psi\rangle^{1}$ and $|\psi\rangle^{9}$, and $\Omega$ is not big enough to induce the off-resonant transitions, then the effective system Hamiltonian in the interaction picture is approximately
 \begin{equation}
 H_{tot}^{int}=\Omega^{'}(|\psi\rangle^{1}\otimes{}^{9}\langle\psi|+|\psi\rangle^{9}\otimes{}^{1}\langle\psi|),
 \end{equation}
 where $\Omega^{'}={}^{9}\langle\psi|\Omega\sigma_{c}^{+}|\psi\rangle^{1}$ is the effective transition element. Start with the initial state $|\psi\rangle^{1}$, the evolution of the system can be described by
 \begin{equation}
 |\Psi(t)\rangle=\cos(\Omega^{'}t)|\psi\rangle^{1}+\sin(\Omega^{'}t)|\psi\rangle^{9}.
 \end{equation}
 Because $|\psi\rangle^{1}$ and $|\psi\rangle^{9}$ describe a 2-site JCHM in localized and delocalized phase, respectively, Eq.\,(7) can be seen as the Rabi
 oscillation between two distinct quantum phases. By choosing $t=\frac{\pi}{4\Omega^{'}}$, the system can be engineered into a Schr\"{o}dinger cat superposition of localized and delocalized phases. Here, the ``dead cat'' corresponds to the resonators are decoupled and the polaritons are frozen in local sites, and the ``living cat''
 corresponds to the resonators are strongly coupled and the polaritons are delocalized through the sites.

 \begin{figure}[!b]
 \includegraphics*[scale=0.39]{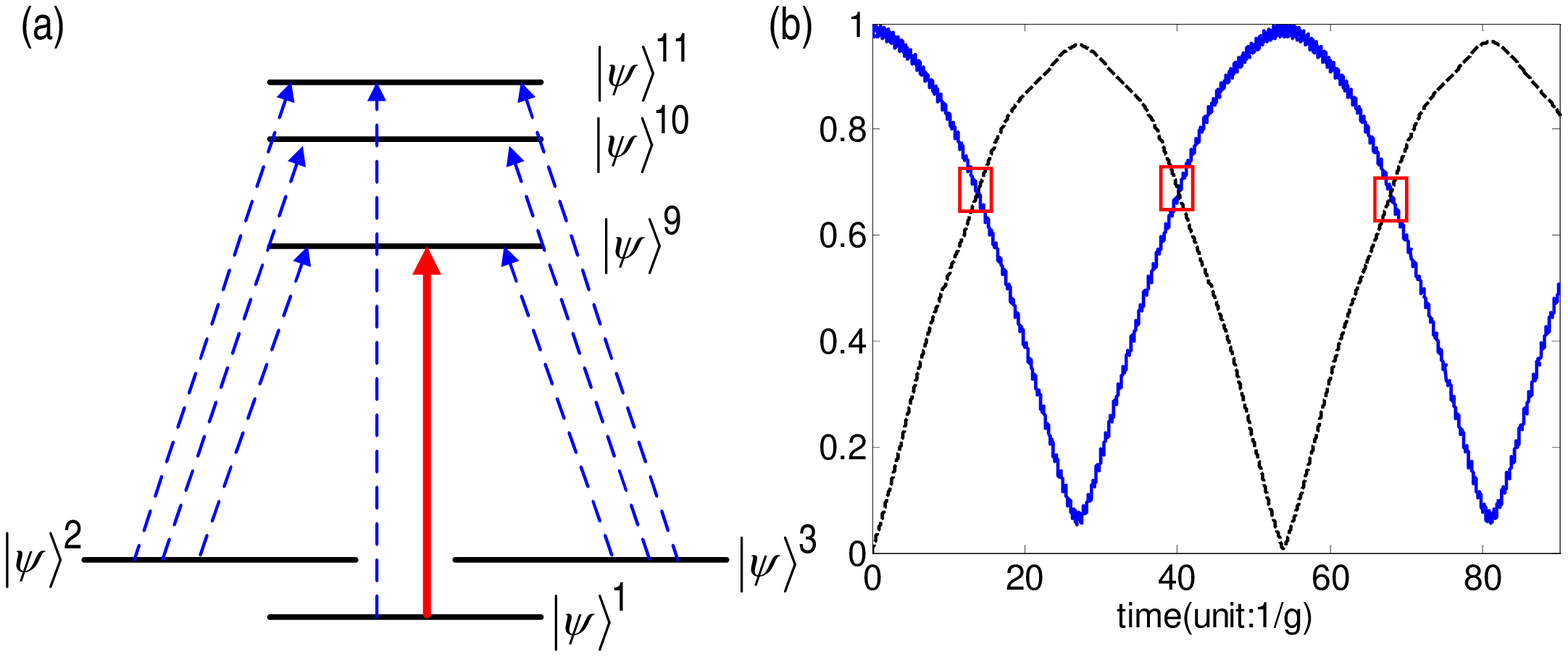}
 \caption{\label{fig4}(Color online) (a) The related nonzero transition elements between the eigenstates when we pump the system from $|\psi\rangle^{1}$ to $|\psi\rangle^{9}$. The solid arrow represents the designed transition, and dashed arrows represent the unwanted off-resonant transitions. (b) The dynamics of the system governed by the exact Hamiltonian $H_{tot}$. The solid and dashed lines represent the time-dependent values of $|{}^{1}\langle\psi|\Psi^{'}(t)\rangle|$ and $|{}^{1}_{e}\langle\phi|\langle e^{c}|\Psi^{'}(t)\rangle|$, respectively. }
 \end{figure}

 For a practical situation, the errors induced by the off-resonant transitions have to be considered. In Fig.\,4(a), we present all the related nonzero transition elements between the eigenstates when we pump the system from $|\psi\rangle^{1}$ to $|\psi\rangle^{9}$. The solid arrow represents the designed transition, and dashed arrows represent the unwanted off-resonant transitions. The minimal off resonance can be obtained as $\delta=min(E_{2},E_{10}-E_{9}-E_{2})$. To suppress the off-resonant transitions, the transition element $\Omega^{'}$ is required to be smaller than $\delta$. In Fig.\,4(b), we show the dynamics of the system governed by the exact Hamiltonian $H_{tot}$ in Eq.\,(5). The parameters we choose are $g_{c}=g$, $\kappa_{0}=0.1g$, $\epsilon=41g$, $\epsilon_{c}=50g$, $w=40g$, $w_{d}=50.2750g$ and $\Omega=0.07g$, which yield $\delta=0.2151g$ and $\Omega^{'}=0.0495g$ \cite{reve16}. The solid and dashed lines represent the time-dependent values of $|{}^{1}\langle\psi|\Psi^{'}(t)\rangle|$ and $|{}^{1}_{e}\langle\phi|\langle e^{c}|\Psi^{'}(t)\rangle|$, respectively, where $|\Psi^{'}(t)\rangle$ is the exact system state obtained by numerically integrating $H_{tot}$. At the time instants marked by small rectangles, the system is engineered into the equal superposition of $|g^{c}\rangle|1-\rangle_{1}|1-\rangle_{2}$ and $|e^{c}\rangle|\phi\rangle_{e}^{1}$.

 In principle, this scheme can be generalized to the more complicated cases (e.\,g.\,connecting a larger number of resonators and injecting more polaritons). However, some problems will arise when we deal with a larger system. First, the complicated system has a very large Hilbert space and a very dense energy spectrum, so it is difficult to identify the designed transitions using spectroscopic techniques. Second, the increase in the number of polaritons will lead to a shorter coherence time of the system, so it is difficult to complete the wanted operations before dissipations occur.

 \emph{Controllable interbit coupling with infinite on/off ratio.---} Superconducting circuits are promising candidates for constructing quantum bits because of their potential suitability for large-scale quantum computation \cite{reve17}. Usually, the coupling and decoupling of the superconducting qubits are implemented by tuning their frequencies in and out of resonance, respectively \cite{reve12,reve18}. The residual interaction that exists when the qubits are detuned from each other, however, limits the accuracy of these proposals. A controllable coupling mechanism, which has infinite on/off ratio, is desirable. Here, we will show that, the circuit proposed in this letter can be used to achieve this goal. Let us use the zero-polariton state and lower 1-polariton state of resonator $i$ to represent the two states of logical qubit $i$, i.\,e.\,, $|0\rangle_{L}^{i}\equiv|0-\rangle_{i}=|0,g\rangle_{i}$, and $|1\rangle_{L}^{i}\equiv|1-\rangle_{i}=\cos\theta_{1}|0,e\rangle_{i}-\sin\theta_{1}|1,g\rangle_{i}$, where $\tan\theta_{1}=\frac{\Delta}{2g}+\sqrt{(\frac{\Delta}{2g})^{2}+1}$. We make the knob qubit $q_{c}$ always stay in its ground state $|g^{c}\rangle$, then the two resonators are coupled by a strength $J=\kappa_{0}-\frac{g_{c}^{2}}{\Delta_{c}}$. To decouple the logical qubits, we tune $\Delta_{c}$ to an appropriate value so that $\kappa_{0}=\frac{g_{c}^{2}}{\Delta_{c}}$ and $J=0$. To switch on the coupling, we tune $\Delta_{c}$ to another value so that $J=\kappa_{0}-\frac{g_{c}^{2}}{\Delta_{c}}\neq0$ but $J$ is much smaller than the effective repulsive energy $u_{r}$. In this case, if one local resonator has a polariton in it, the strong photon blockade effect will prevent a second polariton from entering it. Finally, one can easily get the state evolution of the system:
 \begin{equation}
 |0\rangle_{L}^{1}|0\rangle_{L}^{2}\rightarrow|0\rangle_{L}^{1}|0\rangle_{L}^{2},\qquad|1\rangle_{L}^{1}|1\rangle_{L}^{2}\rightarrow|1\rangle_{L}^{1}|1\rangle_{L}^{2},\nonumber
 \end{equation}
 \begin{equation}
 |0\rangle_{L}^{1}|1\rangle_{L}^{2}\rightarrow\cos(J^{'}t)|0\rangle_{L}^{1}|1\rangle_{L}^{2}-i\sin(J^{'}t)|1\rangle_{L}^{1}|0\rangle_{L}^{2},\nonumber
 \end{equation}
 \begin{equation}
 |1\rangle_{L}^{1}|0\rangle_{L}^{2}\rightarrow\cos(J^{'}t)|1\rangle_{L}^{1}|0\rangle_{L}^{2}-i\sin(J^{'}t)|0\rangle_{L}^{1}|1\rangle_{L}^{2},
 \end{equation}
 where $J^{'}=(\kappa_{0}-\frac{g_{c}^{2}}{\Delta_{c}})\sin^{2}\theta_{1}$ is the effective polariton hopping rate. By choosing $t=\frac{\pi}{4J^{'}}$, we can realize the $\sqrt{i\mbox{\scriptsize {SWAP}}}$ gate of two logical qubits. This architecture is scalable to a large number of logical qubits and may be specially suitable for implementing one-way quantum computation \cite{reve19}.

 \emph{Conclusion.---} In this letter, we propose a method to engineer two microwave resonators into a quantum superposition of being decoupled and strongly coupled (correlated with a knob qubit in ground state and excited state). Using a variant of 2-site Jaynes-Cummings-Hubbard model, we generate entanglement between the distinct quantum phases of the injected polaritons and the internal states of the knob qubit. This architecture can also be used to solve the annoying on/off ratio problem of conventional proposals for scalable quantum computation. Our proposed circuit may play an important role in quantum engineering of novel states of microwave photons and quantum information processing with superconducting qubits.

 This work was supported by FANEDD (Grant No.\,200524), NCET (Grant No.\,06-0920), and NSFC (Grant No.\,11074307).


\begin{thebibliography}{20}
\bibitem[16]{reve16} The slight discrepance between $w_{d}$ and $E_{9}-E_{1}$ results from the higher order
                        terms we have neglected in $H^{eff}$.
\end{thebibliography}
\end{document}